\documentstyle[10pt]{article}
\input{psfig.sty}
\begin{document}

\title{{\bf The Possibility of Thermal Instability in Early-Type Stars Due to
Alfv\'en Waves }}
\author{D.R. Gon\c calves\thanks{%
denise@plasma.iagusp.usp.br}, V. Jatenco-Pereira and R. Opher}
\date{Instituto Astron\^omico e Geof\'\i sico - USP - Av. Miguel Stefano 4200,
S\~ao Paulo SP, 04301-904, Brazil}
\maketitle

\begin{abstract}
It was shown by dos Santos et al. the importance of Alfv\'en waves to
explain the winds of Wolf-Rayet stars. We investigate here the possible
importance of Alfv\'en waves in the creation of inhomogeneities in the winds
of early-type stars. The observed infrared emission (at the base of the
wind) of early-type stars is often larger than expected. The clumping
explains this characteristic in the wind, increasing the mean density and
hence the emission measure, making possible to understand the observed
infrared, as well as the observed enhancement in the blue wing of the 
$H_\alpha$ line. In this study, we investigate the formation of these clumps
a via thermal instability. The heat-loss function used, $H(T,n)$, includes
physical processes such as: emission of (continuous and line) recombination
radiation; resonance line emission excited by electron collisions; thermal
bremsstrahlung; Compton heating and cooling; and damping of Alfv\'en waves.
As a result of this heat-loss function we show the existence of two stable
equilibrium regions. The stable equilibrium region at high temperature is
the diffuse medium and at low temperature the clumps. Using this reasonable
heat-loss function, we show that the two stable equilibrium regions can
coexist over a narrow range of pressures describing the diffuse medium and
the clumps.
\end{abstract}

\section{Introduction}

As demonstrated first by Lucy \& Solomon (1970), the radiative momentum
absorbed by UV spectral lines is able to initiate stellar winds, since the
radiative line acceleration exceeds the gradient by a large factor. The
first model to derive mass--loss rates ($\dot{M}$) and flow speeds in good
agreement with observations was that of Castor, Abbott \& Klein (1975)
(CAK). One of the major difficulties presented by the radiation driven wind
theory is the momentum problem in WR stars, which can be described using the
ratio $\eta =({\dot{M}}v_\infty )/(L_{*}/c)$, where $v_\infty $ is the
terminal velocity and $L_{*}$ the star luminosity. Barlow et al. (1981)
found that in WR stars $\eta $ ranges from 3 to 30. This means that there is
about an order of magnitude more momentum in the wind than in the radiation
field. It was assumed that every stellar photon transfers its momentum, $%
h\nu /c$, only once (single scattering), but even with multiple scattering
of the photons one obtains ${\dot{M}}v_\infty >5L_{*}/c$. To get around the
momentum problem, one cannot simply appeal to a larger luminosity, because
the values that are used cannot be near the Eddington limit (Cassinelli \&
van der Hucht 1987).

Following the suggestion that there may be appreciable magnetic fields in WR
stars larger than 1000 G (Mahesvaran \& Cassinelli 1988; Poe et al. 1989),
it was suggested that the wind in a WRN5 star, for instance, can be driven
by Alfv\'{e}n waves (see Hartmann \& Cassinelli 1981). They assumed B=20,000
G and a mechanical flux of Alfv\'{e}n waves of $\Phi _w=1.1\times 10^{14}\;%
{\rm erg/cm}^2{\rm s}$ (this work did not take into account the contribution
of the radiation pressure on the lines).

As implied by the work of Willis (1991), an additional mechanism to
radiation pressure may be required to initiate the high WR mass--loss,
although thereafter the winds may be radiatively accelerated. In this
context, dos Santos et al. (1993a,b) proposed a model for mass--loss in WR
stars, where both a flux of Alfv\'{e}n waves and radiation pressure are
considered. The model is a fusion of the Alfv\'{e}n wave wind model of
Jatenco--Pereira \& Opher (1989a,b) and the radiation pressure CAK model. In
the model an effective escape velocity is used, which takes into account the
CAK power index expressing the effect of all lines, possible nonsolar
abundances, and the finite size of the star disk. Their work indicates that
Alfv\'{e}n waves, acting jointly with radiation pressure, provide the
necessary energy and momentum for the wind, with reasonable Alfv\'{e}n
fluxes and magnetic fields.

Early-type stars show superionization lines O VI, N V, $H_{\alpha}$ and
X-rays, that cannot be explained by the high temperature star. Applying the
coronal zone model to the winds of early-type stars, Cassinelli \& Olson
(1979) derived the ionization conditions expected in the wind of $\zeta$
Pup. The results of this study explain very well the persistence to low
effective temperatures of the strong lines of O VI, N V, C IV and Si IV.

Since Abbott et al. (1984), one knows that the observed IR emission is often
larger than expected from a homogeneous wind. From that time it was pointed
out that the clumping in the wind, increases the mean density and hence the
emission measure. Clumping can also explain an observed enhancement in the
blue wing of the $H_\alpha $ line. The narrow absorption components are
likely to be direct manifestations of dense clumps. Now, the existence of
these clumps is largely known in many individual hot stars, and clumping may
be important in all hot stars with winds (Hillier 1991; Robert 1994; Moffat
\& Robert 1994; Massa et al. 1995; Brown et al. 1995; Moffat 1994,1996;
Eversberg et al. 1996).

In principle, it can be said that the series of papers by Owocki, Rybicki
and Castor (Owocki \& Rybicki 1984, 1985, 1986, 1991; Owocki et al. 1988;
Rybicki \& Owocki 1990), which contains numerical hydrodynamic calculations,
show that the radiation driven winds are violently unstable and that the
consequent shocks can explain the X-ray emission of early-type stars,
moreover the clumping explains the infrared emission excess and the
formation of narrow absorption components. These models {\it qualitatively}
explain the hot stars wind structure, and the role of Alfv\'{e}n wave
damping in these winds is an open question, as noted by these authors. In
this sense, our model is an attempt to explain the inhomogeneities of hot
star winds using thermal instability in the presence of Alfvenic heating.

The propagation and transmission of magnetohydrodynamic waves through
stellar atmospheres and winds has attracted considerable interest, because
of its relevance to the questions of chromospheric and coronal heating and
wind acceleration (Leer et al. 1982). In order to accelerate a wind
efficiently, the waves must be able to propagate without much reflection or
attenuation up to the sonic point, because any addition of momentum below
that point essentially goes to increase the mass flux but not the asymptotic
wind speed (Leer et al. 1982). On the other hand, as commented by Velli
(1993), waves reflected and/or having a nonlinear evolution in the lower
atmospheric layers, contribute to the nonradiative heating through turbulent
decay.

We study here a mechanism to form condensations in the base of early-type
star winds. The basis for this approach is the work of dos Santos et al.
(1993a,b) -- a wind acceleration model for Wolf-Rayet stars where Alfv\'{e}n
waves act jointly with the radiation pressure. Considering the above
scenario, with special attention to the clumping features, our intent is to
study the relevance of a flux of Alfv\'{e}n waves in the hot medium present
in the base of WR winds, in order to understand the formation of the
clumping features. As the magnetic field is more effective in the base of
the wind (see dos Santos et al. 1993a), we have there a wave flux, resulting
in a model that could explain clumps in this region. We consider a thin hot
corona atmosphere, which corresponds to the X-rays observed, with a
temperature of $\sim 10^7{\rm K}$, and density of $\sim 2\times 10^{13}{\rm %
cm}^{-3}$ (see van der Hucht 1992). In principle, the thermal instability
process presents, as a result, the clumps observed, and we call these clumps
the cool atmosphere ($\sim 10^4{\rm K}$ and $\sim 10^3$ times denser than
the hot medium) (for instance, Brown et al. 1995). Our final goal is to
demonstrate the stability of the base of the wind that has ``cool'' ($\sim
10^4{\rm K}$) clouds and ``hot'' ($\sim 10^7{\rm K}$) intercloud medium
coexisting at the same pressure.

\section{Physical Mechanisms}

In general, astronomical objects are formed by self-gra\-vi\-ta\-ti\-on.
However, some objects cannot be explained by this process. For these objects
the gravitational energy is smaller than the internal energy. In these
cases, it is assumed that the internal pressure is balanced by the pressure
of the external medium. These objects (that cannot be explained by
self-gravitation) are formed from the medium by some kind of condensation
process not involving gravitation. Parker (1953) argued that, if the thermal
equilibrium of the medium is a balance between energy gains and radiative
losses, instability results if, near equilibrium, the losses increase with
decreasing temperature. Then, a cooler-than-average region cools more
effectively than its surroundings, and its temperature rapidly drops below
the initial equilibrium value.

Following Lucy \& Solomon (1970), for a given ionization potential, $\chi
_r=h\nu _r$, the photoionization rate, $\Gamma _r$ is a function only of the
radiation temperature, $T_r$; the collisional rate, $\Gamma _c$, on the
other hand, is determined by the electron density, $n_e$, and the electron
temperature, $T_e$. The ratio of the two rates is 
\[
\frac{\Gamma _r}{\Gamma _c}=\frac{\left[ \int_{\nu _r}^\infty \frac{4\pi
B_\nu (T_r)}{h\nu }a_\nu d{\nu }\right] }{n_e\langle \sigma _rv_e\rangle }%
\;\;\;,
\]
\noindent
here $\sigma _r$ is the cross--section for collisional ionizations, $v_e$
the electron velocity, and $a_\nu $ the photoionization cross--section. B$%
\ddot{{\rm o}}$hm (1960) has given approximations for these quantities from
which one derives 
\[
\frac{\Gamma _r}{\Gamma _c}\simeq 6\times 10^{10}\frac{T_e^{1/2}\chi _r^3}{%
n_e}\;\;\;,
\]
\noindent where $T_r=T_e$, $h\nu _r/kT_r\geq 1$, approximately, and the
units of $\chi $ are electron volts.

Applying these results to the ionization of C III ions, for instance, $\chi
_r=47.9$ {\rm eV} with $\log T_{eff}=4.5$. Taking $T_e$ (in the hot
atmosphere) $\approx 10^3T_{eff}$, and $n_e=2\times 10^{13}{\rm cm}^{-3}$,
one obtains ${\frac{\Gamma _r}{\Gamma _c}}=3\times 10^5$ (for lower density
the ratio is even higher), so that collisional ionization may be completely
neglected.

Considering a thermal instability in an isobaric regime (e.g., Field 1965)
(internal pressure balanced by the external pressure), we looked for a set
of physical parameters that, at equilibrium $[H(T,n)=0]$, show three
equilibrium regions: one stable region representing the diffuse medium; one
unstable region; and another stable region representing the condensations.
The energy gains considered are: heating by photoionization-recombination, $%
H_r$; Compton heating, $H_c$; and Alfv\'{e}n wave heating, $H_A$. These
gains are balanced by the following radiative loss processes: cooling via
thermal bremsstrahlung, $H_b$; inverse Compton cooling (this term is
computed jointly with $H_c$); and collisional excitation followed by
resonance line emission, $H_{rl}$.

\subsection{Bremsstrahlung Losses}

The total amount of energy radiated in free-free transitions, per ${\rm cm}%
^3 $ per ${\rm sec}$, in the case of a Maxwellian distribution of
velocities, is

\[
H_b=-\left( \frac{2\pi kT}{3m_e}\right) ^{1/2}\frac{2^5\pi e^6}{3hm_ec^3}%
Z^2n_en_i{\bar{g}}_{ff} 
\]

\[
H_b=-1.42\times 10^{-27}Z^2n_in_eT^{1/2}{\bar{g}}_{ff} 
\]

\begin{equation}
H_b=-\lambda _bT^{1/2}n^2\;.
\end{equation}

\noindent  The quantity ${\bar{g}}_{ff}$ appearing above is a correction
factor required for precise results. Its value is generally about unity
(Spitzer 1978), and $\lambda _b=2.4\times 10^{-27}$. Hereafter $n_e\equiv n$%
, the number density.

\subsection{Resonance Line Emission}

Raymond et al. (1976) calculated a radiative cooling coefficient for a low
density gas, optically thin, with cosmic abundances, between the
temperatures of $10^4{\rm K}$ and $10^8{\rm K}$, which we adopt in this
work. A good fit to radiative losses, in this temperature range, due to
electron excitation of resonance transitions in common metal ions (${\rm %
erg/cm}^3{\rm s}$) is (Raymond et al. 1976; Mathews \& Doane 1990),

\begin{equation}
H_{rl}\;=\;-\;\frac{a\;T^p}{1\;+\;b\;T^q}\;n^2\;\;,
\end{equation}

\noindent  with $a = 1.53 \times 10^{-27}$, $b = 1.25 \times 10^{-9}$, $p =
1.2$, and $q = 1.85$.

\subsection{Photoionization-Recombination Heating}

An approximate equation to express the residual heating due to radiative
ionization followed by recombination, in ${\rm erg/cm}^3{\rm s}$, is

\begin{equation}
H_r\;=\;\alpha _B\;(T){\rm max}\;\left[ 0,\langle h\nu \rangle _i-h\nu
_o-f3k_BT/2\right] \;n^2\;,
\end{equation}

\noindent  where $\langle h\nu \rangle_i$, the mean energy of ionizing
photons, is

\[
\langle h\nu \rangle _i=\frac{\sum n_jh\nu _j}{\sum n_j}=\frac{%
n_HkT_H+n_LkT_L}{n_H+n_L}\;, 
\]

\noindent  $n_L$ and $n_H$ are the numerical densities of photons of the low
and high temperature regions, respectively, given by

\[
n_L=\frac{\sigma T_L^4}{k_BT_Lc}\;{\rm with}\;T_L=T_{*}\;, 
\]
\noindent  and 
\[
n_H=\frac{[H_{rl}(T_H,n_H)+H_b(T_H,n_H)]tR_{*}}{kT_Hc}\;, 
\]

\noindent  where $\alpha _B=2.60\times 10^{-13}$ $(T/10^4)^{-0.8}${\rm \ }$%
cm^3${\rm \ }$s^{-1}$ is the recombination coefficient, and $h\nu _o$ the
ionization potential of hydrogen. In the above equations $n_H$ and $T_H$ are
the density and temperature of a high temperature region, $n_L$ and $T_L$
are the same for a low temperature region, and $tR_{*}$ is the region
thickness. Each recombination results in a loss of energy $f3k_BT/2$ from
the thermal energy of the plasma, with $f\approx 0.43$ (Mathews \& Doane
1990).

The recombination expression, eq.(3), is an approximated one. Equation (3)
states that we have recombination only when the average energy of the
photons, $\langle h\nu \rangle _i$, is sufficiently high such that
ionization can occur, that is, when $\langle h\nu \rangle _i$ is greater
than $h\nu _0\;+\;f3k_BT/2$ (i.e., the sum of the ionization potential plus
the average energy of the electron that is liberated).

\subsection{Compton Heating-Cooling}

We have to estimate the number and frequency of the photons acting in the
immediate neighborhood of the star surface and anywhere in the star
atmosphere. Taking into account the interaction between thermal electrons
and the radiation field, photons with lower frequency, come from a cooler
optically thick region, the stellar continuum. Their flux is $\sigma {T_{*}}%
^4$, and then, for these photons, we have 
\[
\frac{L_{bL}}{4\pi R{_L}^2}=\sigma {T_{*}}^4\;\;\;. 
\]
\noindent On the other hand, a hot region of thickness $tR_{*}$ (optically
thin), at temperature $T_H$ causes heating in the medium via, principally,
thermal bremsstrahlung and resonance line emission [$H_b(T,n)$ and $%
H_{rl}(T,n)$]. Hence, for these photons, 
\[
\frac{L_{bH}}{4\pi R_H^2}=\left[ H_b(T_H,n_H)+H_{rl}(T_H,n_H)\right]
tR_{*}\;\;\;. 
\]
\noindent The complete expression for Compton heating and cooling is then

\begin{equation}
H_c\;=\;\frac{4\;k_B\;\sigma _T}{m\;c}\frac nc\left\{ \left[ (T_H-T)\frac{%
L_{bH}}{4\pi R_H^2}\right] +\left[ (T_L-T)\frac{L_{bL}}{4\pi {R_L}^2}\right]
\right\} \;\;,
\end{equation}

\noindent  which is similar to the expression usually adopted, for instance
by Mathews \& Doane (1990). In the above equations $k_B$ is the Boltzmann
constant, $\sigma _T$ the Thompson cross section, $\sigma $ the
Stefan-Boltzmann constant, $n$ the number density, $m$ the electron mass, $T$
the temperature, $L_b$ the bolometric luminosity, $c$ the light speed, $%
T_{*} $ the stellar temperature and $R_{*}$ the stellar radius.

\section{Damping and hea\-ting from Alf\-v\'{e}n wa\-ves}

Alfv\'{e}n waves in a early-type star, whose winds are primarily radiatively
driven, are subject to damping (or amplification) as described, for example,
by MacGregor (1996). In this case, the dispersion relation for Alfv\'{e}n
waves in a radiatively driven wind, is $k^2v_A^2=\omega ^2-i\omega \omega _0$
(instead of $k^2v_A^2=\omega ^2$), where 
\[
\omega _0=\frac{2\pi \kappa _L\nu _L}{3c^2}\frac{dI_{*}(\nu )}{d\nu }\qquad
, 
\]

\noindent $I_{*}(\nu )$ is the intensity of the photospheric radiation
field, $\nu _L$ is a line rest frequency and $\kappa _L$ is the line mass
absorption coefficient. If $dI_{*}(\nu )/d\nu >0$ then the Alfv\'{e}n wave
is amplified, while if $dI_{*}(\nu )/d\nu <0$ the Alfv\'{e}n wave is damped.
Although, as MacGregor (1996) noted, ``the presence of such radiatively
modified Alfv\'{e}n waves in the flow has yet to be explored''. We apply in
the present investigation, the dampings described below with their heatings.

The damping mechanisms that we assume here were used before in many
astrophysical objects: protostellar, late-type stars and solar winds \ \
(Jatenco-Pereira \& Opher\- 1989a,b); galactic and extragalactic jets (Opher
\& Pereira 1986; Gon\c {c}alves et al. 1993b); early-type stars (dos Santos
et al. 1993a,b); broad line regions of quasars (Gon\c {c}alves et al. 1993a,
1996); cooling flows of galaxy clusters (Fria\c {c}a et al. 1997) and others.

\subsection{Nonlinear damping}

Parallel Alfv\'{e}n waves are purely transverse and there is no important
linear damping. The damping that does occur is not linear and it arises from
a beat wave (two circularly polarized parallel propagating waves) which
contains a longitudinal field component and a longitudinal gradient in the
magnetic field. This results in a nonlinear damping of both electrostatic
and magnetostatic components i.e. transient time damping.

V$\ddot{{\rm o}}$lk \& Cesarsky (1982) derived an equation that represents
the unsaturated Landau damping, in the case of nonlinear two-wave
interaction, that can be written as 
\[
\;\;\Gamma (k)=\frac 14\sqrt{\frac \pi 2}\xi \bar{k}v_s\Im \;, 
\]

\noindent where $\Im $ is the energy density in waves normalized to the
ambient magnetic energy density, $B_0^2/8\pi $ (Lagage \& Cesarsky 1983).
Using $\Im ={\frac{\rho \langle \delta v^2\rangle }{B_0^2/8\pi }}$ and $\bar{%
k}={\frac \varpi {v_A}}$, we obtain:

\begin{equation}
\Gamma _{nl}={\frac 14}{\ \sqrt{\frac \pi 2}}\xi \bar{w}\left( \frac{v_s}{v_A%
}\right) \frac{\rho \langle \delta v^2\rangle }{B_0^2/8\pi }\;\;\;,
\end{equation}

\noindent  where $\xi =5-10$ and $v_s$ is the sound velocity.

\subsection{Turbulent damping}

There are strong evidences favoring anisotropic, supersonic and compressible
turbulence in WR winds. Since all WR stars observed intensively so far do
behave similarly, and WR stars are extreme manifestations of winds in hot
luminous stars, it is possible or even likely that all hot-star winds show
the same basic phenomenon (Moffat et al. 1994 and references therein). A
necessary (but not sufficient) condition that one is dealing with turbulence
is that the Reynolds number be $Re>>1$. For an expanding wind, with $v_w(r)$
the expansion speed and $r$ the distance from the star, one has: 
\[
Re\approx \frac{rv_w(r)}{\nu _{thermal}}\approx \frac r{l_{mfp}}\frac{v_w(r)%
}{v_s}\;\;\;, 
\]
\noindent  where $\nu $ is the viscosity, $l_{mfp}$ the mean free path of
the average particle in the medium. Thus, with typical WR wind values, where
the observed lines form, $Re$ is much higher than $1$, so turbulence is
likely to exist if there is a driving force.

Hollweg (1986) considered a new hypothesis for the nonlinear wave
dissipation of Alfv\'{e}n waves. The hypothesis is that the wave dissipates
via turbulent cascade, or, this hypothesis concerns the consequences of the
Alfv\'{e}n wave dissipation in terms of wave-particle interactions, where
the required power at high frequencies is presumably supplied via turbulent
cascade. Then, exploiting the similarity of $P_{{\rm B}}\propto k^{-5/3}$
and Kolmogorov turbulence in ordinary fluids, the plasma volumetric heating
rate associated with the cascade is given by: 
\begin{equation}
E_H=\frac{\rho \langle \delta v^2\rangle ^{3/2}}{L_{{\rm corr}}}\frac {}{},
\end{equation}

\noindent  where $\rho $ is the mass density, $\langle \delta v^2\rangle $
is the velocity variance associated with the wave field, and $L_{{\rm corr}}$
is a measure of the transverse correlation length. A subhypothesis is that
the correlation length scales as the distance between magnetic field lines, 
\[
L_{{\rm corr}}\propto B^{-1/2}\frac {}{}. 
\]
\noindent In spite of $L_{{\rm corr}}$ being a free parameter, the model
comes close to the notion of a Kolmogorov-like cascade to small scales. The
waves themselves are here regarded as the source of the heating. In this
case $\langle \delta v^2\rangle $ only includes the power associated with
Alfvenic fluctuations. Similarly, $L_{corr}$ concerns the correlation length
of the Alfvenic fluctuations. Finally, in terms of damping length, we have

\begin{equation}
L_t=L_{{\rm corr}}\;v_A{\langle \delta v^2\rangle }^{-1/2}\frac {}{},
\end{equation}

\noindent  (Hollweg 1986, 1987).

\subsection{Alfv\'en wave heating}

Data of the last 10 years show us that early--type stars can be separated in
two groups: magnetic stars, with surface strengths of a dipole or quadrupole
magnetic field of $B\approx n(10^2-10^3){\rm G}$, $n=2,3,...7$; and normal
stars, with $B\approx 1-100\;{\rm G}$. The magnetic field strength increases
towards the center of the star and in the core is $\approx (0.1-10)\times
10^6{\rm G}$, depending on the stellar mass (Dudorov 1994; Bohlender 1994).
The origin of these fields is an open question, and two theories compete to
explain it: dynamo and fossil theories (Moss 1994).

Consider now a collapsing cloud. For the collapsing cloud the cross
sectional area perpendicular to a magnetic field, $A$, is $\propto \rho ^{-{%
\frac 23}}$ and $B$ $\propto A^{-1}\propto \rho ^{{\frac 23}}$, where $\rho $
is the mass density of the gas and $B$ is the magnetic field. The damping
length in each case is $L=v_A/\Gamma $ (i.e., the ratio between the
Alfv\'{e}n velocity and the damping rate). Knowing that $v_A=B/\sqrt{4\pi
\rho },\;\;\rho \langle \delta v^2\rangle \;\propto \;\Phi _w/v_A$, where $%
\Phi _w$ is the wave flux and $\Phi _w\;\propto \;\rho ^{2/3}$, we write the
nonlinear Alfvenic heating as:

\begin{equation}
H_{nl}\;=\;\Phi _w/L_{nl}\;=\;\frac{\Phi _w}{v_A}\Gamma _{nl}\;\propto \;%
\frac{\Phi _w}{v_A^2}\frac{\rho \langle \delta v^2\rangle }{B^2}\;\propto
\;\rho ^{-1/2}\;\;\;;
\end{equation}

\noindent and for the turbulent Alfvenic heating,

\begin{equation}
H_t=\Phi _w/L_t\;\propto \;\Phi _wB^{1/2}\frac{{\langle \delta v^2\rangle }%
^{1/2}}{v_A}\;\propto \;\rho ^{7/12}\;\;\;;
\end{equation}

\noindent following equations (5) and (7).

The sum of the contributions from Compton and inverse Compton,
photoionization-recombination, bremsstrahlung and resonance line emission,
is about $10^{-22}n^2$. We are adopting $n_H$ (the density of the hot
atmosphere)$=2\times 10^{13}{\rm cm}^{-3}$, resulting $10^{-22}n^2\approx
4\times 10^4{\rm erg/cm}^3{\rm s}$. We then normalize the Alfvenic heatings
using $F_{nl},F_t=10^2-10^5{\rm erg/cm}^3{\rm s}$. So,

\begin{equation}
H_{nl}=F_{nl}\left( \frac n{2.0\times 10^{13}}\right) ^{-{\frac 12}}
\end{equation}

\noindent and

\begin{equation}
H_t=F_t\left( \frac n{2.0\times 10^{13}}\right) ^{\frac 7{12}}\;\;\;.
\end{equation}

\subsection{The overall heating/cooling behavior}

In order to make clear the relevance of each heating/cooling process in the
overall balance we plot, in Figure 1, the module of the heating or cooling
due to: resonance line emission; photo\-io\-ni\-zation-recombination;
thermal bremsstrahlung; Alfvenic turbulent and nonlinear heatings; and
Compton interactions, as a function of temperature, in units of ${\rm erg/cm}%
^3{\rm \;s}$.

The first characteristic we note from Figure 1 is the fact that resonance
line emission is the most important cooling. In fact it dominates over all
the other processes in this range of temperature. Another aspect is that we
are using only the contribution of the photoionization-recombination
processes that produces heating. This mechanism is not considered at
temperatures higher than $\log T\simeq 6.3$. At these high temperatures it
appears as cooling (see also eq. (3)).

From the physics of Alfv\'{e}n wave heating, it is clearly not temperature
dependent (eqs. (10) and (11)), beyond a dependence on the density which
implicitly scales inversely to temperature, keeping P fixed. Then, as
Alfv\'{e}n heating is proportional to $n^\alpha $ ($\alpha =-1/2,7/12$), at
a given pressure, we have this heating proportional to $T^{-\alpha }$, as
can be seen from Figure 1.

\section{Results}

The complete heating--cooling function, $H(T,n)$, including the physical
processes discussed above, is: 
\[
H(T,n)=-\lambda _bT^{{\frac 12}}n^2-\frac{aT^p}{1+bT^q}n^2+\frac{4k_B\sigma
_Tn}{mc^2}\times 
\]

\[
\left\{ (T_H-T)\left[ \frac {}{}H_b(T_H,n_H)+H_{rl}(T_H,n_H)\right]
tR_{*}+(T_L-T)\frac {}{}\sigma T_{*}^4\right\} + 
\]
\begin{equation}
\alpha _B(T){\rm max}\left[ 0,\langle h\nu \rangle _i-h\nu _o-\frac{f3k_BT}%
2\right] n^2\ +\ H_A\ \ \ ,
\end{equation}
\ \ \ \ \ \ \ \ \ \ 

\noindent with $H_A$ assuming the form of $H_{nl}$ and $H_t$, given by
equations (10) and (11). All the constants in (12) are in $cgs$ units.

Figures 2 to 5 show the balance between energetic gains and radiative losses
(from equation (12)), i.e., the equilibrium of $H(T,n)(\approx 0)$ in a $%
\log P-\log T$ diagram. For this calculation we assume $n_H=2.0\times 10^{13}%
{\rm cm}^{-3}$, $T_H=1.0\times 10^7{\rm K}$, $T_L=T_{*}=3.0\times 10^4{\rm K}
$; $R_{*}=12R_{\odot }$ and $t=0.2$.

As we are forming clouds, via thermal instability, from the hot atmosphere ($%
n_H=2.0\times 10^{13}{\rm cm}^{-3}$ and $T_H=1.0\times 10^7{\rm K}$), we
performed calculations in order to find, for each temperature, the density
that corresponds to the balance ($H(n,T)\approx 0$). From the isobaric
instability criterion, we need the clouds and the hot medium coexisting at
the same pressure. In the hot atmosphere the pressure is $%
P_H=2n_HkT_H\approx 5.5\times 10^4{\rm dyn/cm}^2{\rm s}$. We then want to
verify the possibility of forming clouds ($\approx 10^4{\rm K}$, cool and
dense atmosphere) at pressures near the characteristic pressure of the hot
medium.

In Figures 2 -- 5 we assume a single value of $F_A$ for the two Alfvenic
heating terms, nonlinear ($F_{nl}$) and turbulent ($F_t$). In Figure 2, we
have the equilibrium diagram in terms of pressure and temperature adopting $%
F_A=10^2$. This figure shows us the coexistence of the two-phase equilibrium
at pressures lower than the characteristic pressure of the hot atmosphere ($%
{\rm \log }P_H=4.74$). For this value of $F_A$, the nonlinear heating (in
the case in which the Alfvenic heating contribution to the overall
heating-cooling function is the nonlinear one), as well as the turbulent
heating, are insufficient to permit the coexistence of the hot and cool
atmospheres, at appropriate pressures. Analyzing Figure 3 ($F_A=5\times 10^3$%
), we observe that both cases (nonlinear and turbulent) are satisfactory, in
order to permit cloud formation via thermal instability, since they reach
the pressure desired. Finally, in Figure 4 ($F_A=10^4$) and Figure 5 ($%
F_A=10^5$), the cool and hot stable solution can be found at the pressure
desired (the reference pressure and higher pressures) in the two cases. In
addition, it is clear that turbulent Alfvenic heating does a better job than
nonlinear heating, since the pressure range of the coexistence of both
stable equilibria is bigger (see bottom panel in Figures 4 and 5) in the
case of this Alfvenic heating.

\section{Discussion and Conclusions}

Our results can be discussed in terms of the efficacy of the Alfvenic
heatings in forming condensations near the surface of WR stars due to
thermal instability. This efficiency is included in the scaling factor $F_A$%
, that is equal $F_{nl}$ or $F_t$ in Figures 2 - 5.

We consider in this study a heuristic derivation of the expressions for
Alfvenic heatings. The nonlinear and turbulent Alfvenic heatings represent
extreme opposite dependencies of these heatings on density. Comparing this
behavior we have: turbulent heating ($H_t\propto n^{7/12}$) which deposits
more energy when the density is higher; on the other hand, nonlinear heating
($H_{nl}\propto n^{-1/2}$) deposits more energy when the density is lower.
These behaviors also can be seen from Figure 1 in which cooler regions are
denser than hotter ones (that figure was plotted for a fixed pressure ($P_H$%
), then, the nonlinear heating curve is a decreasing function of density,
opposite to the case of turbulent heating). Due to the completely different
behavior of these heatings, the results from each one are very different
(see plots 2 to 5 with the equilibrium solution for our models). Despite the
fact that the Alfvenic heatings do not work in the cool solution, as well as
in the hot one, results with nonlinear heating produce the cool
condensations at about $2.2\times 10^4{\rm K}$, while the results with
turbulent heating show cooler low temperature solutions ($\sim 1.0\times 10^4%
{\rm K)}$. Noting also the way that each process scales with P, at fixed T,
one can understand why the stable hot solution has a narrower range in
pressure than the stable cool solution, in all figures of equilibrium
(figures 2 - 5).

In the work of dos Santos et al. (1993a), the principal emphasis was to
determine the terminal velocity of the Wolf-Rayet star winds, using a model
which had radiation pressure and Alfv\'{e}n waves driving the wind. The
initial Alfv\'{e}n wave flux, $\Phi _w$, required was $\approx 5.6\times
10^{12}{\rm erg/cm}^2{\rm s}$. Using this value for the wave flux, we can
estimate the damping length for the Alfv\'{e}n waves. As in the derivation
of Alfvenic heatings (subsection 3.3), $\Phi _w$ is equal to $H_AL_A$.
Adopting, for example, the maximum pressure in which the stable two-phase
equilibrium exists, in Figure 5b, $P_H\simeq 5\times 10^5\;{\rm dyn/cm}^2$,
the turbulent Alfvenic heating for the formed clouds is $\sim 1.9\times
10^7\;{\rm erg/cm}^3{\rm s}$. Then, 
\[
L_{t(min)}=\frac{\Phi _w}{H_t}\simeq {\frac{5.6\times 10^{12}}{1.9\times 10^7%
}}\simeq 3\times 10^5\;{\rm cm}. 
\]
\noindent  Taking now the minimum value for the pressure in the cloud in
Figure 5b, $P\simeq 3.9\times 10^4{\rm dyn/cm}^2$, the turbulent Alfvenic
heating is $\sim 4\times 10^6{\rm erg/cm}^3{\rm s}$. We then have 
\[
L_{t(max)}=\frac{\Phi _w}{H_t}\simeq {\frac{5.6\times 10^{12}}{4\times 10^6}}%
\simeq 1.4\times 10^6\;{\rm cm}. 
\]
\noindent  These values for the Alfvenic damping lengths can be understood
as limits to the size of the formed clouds, since the cloud diameters ($d_c$%
) must be smaller than the damping lengths $(3\times 10^5\leq d_c\leq
1.4\times 10^6{\rm cm}$), in order to have turbulent damping of the
Alfv\'{e}n waves effective in this cloud formation process.

The Alfv\'{e}n flux adopted here was the necessary value in order to
accelerate the wind to the observed velocity, and obtain the necessary
momentum in the wind, with minimum magnetic field. The ratio between this
flux and the total one, at the star surface is 
\[
\xi =\frac{\Phi _w}{L_{*}/4\pi {R_{*}}^2}\;\;\;. 
\]
\noindent
Assuming typical values for $L_{*}$ ($10^{5.5}L_{\odot }$) and $R_{*}$ (12 $%
R_{\odot }$), we obtain $\xi \approx 0.04$, which means that $\Phi _w$ is
only a few percent of the total stellar flux at the star surface. (This flux
of waves is lower than others used in the literature, for instance, Hartmann
\& Cassinelli 1981.)

It is interesting to estimate also the magnetic field before (in the diffuse
medium, $B_a$) and after (in the clouds, $B_c$) the condensation process.
During the collapse the density increases $10^3$ -- $10^4$ times. Since $%
B\propto \rho ^{2/3}$, the magnetic field increases $10^2$ -- $10^{2.7}$
times during the cloud formation (we are considering the magnetic field
frozen in the plasma). For a pressure of $\sim 5.5\times 10^4{\rm dyn/cm}^2$%
, the maximum magnetic field in the cloud, in order to have the magnetic
field not dominating the pressure, is 
\[
{\frac{B^2}{4\pi }}\leq 5.5\times 10^4\qquad {\rm or}\qquad B_c\leq
8.3\times 10^2\;{\rm G}. 
\]
\noindent  In this way, the ambient magnetic field is then $B_a\leq
(1.65-8.3)\;{\rm G}.$

Our treatment can be understood in the light of the model of dos Santos et
al. (1993a,b) that focuses on the mass loss from WR stars due to radiation
pressure and Alfv\'{e}n waves. As radiation pressure line-driven models have
difficulties in explaining some observational characteristics of these stars
(such as the disagreement between the observational mass loss rates and the
maximum predicted mass loss by radiation pressure), they proposed that a
flux of Alfv\'{e}n waves and radiation pressure act jointly. This fusion of
an Alfv\'{e}n wave wind model (Jatenco-Pereira \& Opher 1989a) and the
radiation pressure CAK model resulted in good agreement with observations.
In the present model we adopted the Alfv\'{e}n wave flux from the model of
dos Santos et al. (1993a,b), and showed that the heating due to the
Alfv\'{e}n wave flux can cause a thermal instability which results in cold
dense clumps coexisting with a hot diffuse gas. It is important to note also
that near the star surface the magnetic force is more efficient than
elsewhere (this is seen, for instance, in Figures 7 - 9 of dos Santos et al.
1993a). The thermal instability processes that we propose occurs just near
the stellar surface.

Another important aspect to discuss here is the competition between the
instabilities acting in the wind. In principle, any thermal instability of
the wind material would have to compete against the intrinsic line-driven
instability of the flow. Then, consider that the cooling time, or thermal
instability time, in the unperturbed medium, is

\[
t_{cool}\approx \frac{k_BT}{n\Delta }\simeq 1.6{\rm s}\qquad ({\rm at}\qquad
P_H=5.5\times 10^4{\rm dyn/cm}^2)\frac {}{}, 
\]

\noindent where $T\equiv T_H$, $k_B$ is the Boltzman constant, and $\Delta $
is the cooling rate dominated by radiative processes ($H_b+H_{rl}$).
Consider also that the dynamical time, or the line-driven instability time,
at the base of the wind (up to one stellar radius), is

\[
t_{dyn}\approx \frac R{\left\langle v\right\rangle }\simeq 1.6\times 10^3%
{\rm s}\frac {}{}, 
\]
\noindent where $R$ is the distance from the ionization source ($\sim
0.1R_{*}$) and $\left\langle v\right\rangle \simeq 5.04\times 10^7{\rm cm/s}$
(Owocki 1994; dos Santos et al.1993a) is the wind velocity in this region.
The above estimates show that the cooling time is much smaller than the
dynamical time, as is necessary in order to have the thermal instability
predominate in this region (see Krolik 1988; Mathews \& Doane 1990).
Moreover, the analysis of what instability is the most relevant for the
thermodynamic of the wind is related to the region in which each instability
operates. For instance, following Owocki (1994), we know that in the
line-driven instability a minor part of the material is actually accelerated
to high speed, for most of the mass the dominant effect is clumping. Diffuse
radiation plays an important role in reducing the line-driven instability,
especially near the wind base. The competition between these two
instabilities may be important far in the wind, but at the base it is not.
In this region thermal instability works alone.

Thinking about the mass quantity involved in this condensation process, we
can also note that if at a given pressure a thermal instability is
indicated, namely matter at a high temperature and a low temperature is
permitted, little can be said about the fraction of the matter that is at
the high or low temperature. For example, for the Crab nebula observations
we have that almost all the matter is found to be contained in the filaments
(Wilson 1971). These filaments are generally attributed to have formed by a
thermal instability (e.g., Gouveia Dal Pino \& Opher 1989).

In spite of the presence of a gradient in the temperature, we did not
consider the effect of thermal conduction in our model. We proceeded in this
way because thermal conduction is extremely reduced in the gas, due to the
presence of the magnetic field.This reduction occurs in the direction
perpendicular to the field lines (Field 1965), and it is so important that
even small magnetic fields are sufficient in eliminating this component of
the thermal conduction (for some applications, see Begelman \& McKee 1990
and McKee \& Begelman 1990).

\bigskip

In conclusion, our principal goal in this study was to explore whether or
not a thermal instability, assisted by Alfvenic heatings, can play a role in
the base of early-star winds. In spite of the fact that a number of
simplifications were adopted in this first investigation, our results limit
the pressure range for the existence of the two-phase equilibrium at the
base of these winds. Our calculations indicate that a thermal instability
may be a viable mechanism to form clumps in the winds of early-type stars.

\vspace{1 true cm}

\noindent  {\bf Acknowledgements} We would like to thank the referees for
helpful suggestions that improved the paper. One of the authors (D.R.G.)
would like to thank the Brazilian agency FAPESP (92/1403-4) for support, and
the other authors (V.J.P. and R.O.) would like to thank the Brazilian agency
CNPq for partial support.

\vspace{1 true cm}

\noindent  {\bf References}

\noindent   Abbott D.C., Telesco C.M., Wolf S.C., 1984, ApJ 279, 225

\noindent   Barlow M.J., Smith L.J., Willis A.J., 1981, MNRAS 196, 101

\noindent   Begelman M.C., McKee C.F., 1990, ApJ 358, 375

\noindent   
\hangindent=0.5 true cm Bohlender D.A., 1994, in: Pulsation, Rotation and
Mass Loss in Early-Type Stars, Symp. IAU 126, p. 155

\noindent   
\hangindent=0.5 true cm B$\ddot{{\rm o}}$hm K.H., 1960, in: Stars and
Stellar Systems, vol 6, J.L. Greenstein (ed.) -- Chicago: University of
Chicago Press, p. 101

\noindent   
\hangindent=0.5 true cm Brown J.C., Richardson L.L, Antokhin I., Robert C.,
Moffat A.F.J., St-Louis N., 1995, A\&A 295, 725

\noindent   Cassinelli J.P., Olson G.L., 1979, ApJ 229, 304

\noindent   
\hangindent=0.5 true cm Cassinelli J.P., van der Hucht K.A., 1987, in:
Instabilities in Early--Type Stars, H. Lamers \& C.M.H. de Loore (eds.),
Dordrecht: Reidel, p. 231

\noindent   Castor J.I., Abbott D.C., Klein R.I., 1975, ApJ 195, 157

\noindent   dos Santos L.C., Jatenco-Pereira V., Opher R., 1993a, ApJ 410,
732

\noindent 
\hangindent=0.5 true cm dos Santos L.C., Jatenco-Pereira V., Opher R.,
1993b, A\&A 270, 345

\noindent   
\hangindent=0.5 true cm Dudorov A.E., 1994, in: Pulsation, Rotation and Mass
Loss in Early-Type Stars, Symp. IAU 126, p. 184

\noindent   
\hangindent=0.5 true cm Eversberg T., L\'epine S., Moffat A.F.J., 1996, in:
Wolf-Rayet Stars in the Framework of Stellar Evolution, $33^{rd}$ Li\`ege
Int. Astroph. Coll.

\noindent   Field G.B., 1965, ApJ 142, 531

\noindent   
\hangindent=0.5 true cm Fria\c {c}a A.C.S., Gon\c {c}alves D.R., Jafelice
L.C., Jatenco-Pereira V., Opher R., 1997, A\&A 324, 449

\noindent   Gon\c calves D.R., Jatenco-Pereira V., Opher R., 1993a, ApJ 414,
57

\noindent   
\hangindent=0.5 true cm Gon\c calves D.R., Jatenco-Pereira V., Opher R.,
1993b, A\&A 279, 351

\noindent  
\hangindent=0.5 true cm Gon\c calves D.R., Jatenco-Pereira V., Opher R.,
1996, ApJ 463, 489

\noindent   Gouveia Dal Pino E.M., Opher R., 1989, MNRAS 240, 573

\noindent   Hartmann L.W., Cassinelli J.P., 1981, BAAS 13, 785

\noindent   Hillier D.J., 1991, A\&A 247, 455

\noindent   Hollweg J.V., 1986, J. Geophys. Res. 91, 4111

\noindent   
\hangindent=0.5 true cm Hollweg J.V., 1987, in Proc. 21st ESLAB Symp. on
Small-Scale Plasma Process (Esa SP - 275), p. 161

\noindent   Jatenco-Pereira V., Opher R., 1989a, MNRAS 236, 1

\noindent   Jatenco-Pereira V., Opher R., 1989b, A\&A 209, 327

\noindent Krolik J.H., 1988, ApJ 325, 148

\noindent   Lagage P.O., Cesarsky C.J., 1983, A\&A 125, 249

\noindent   Leer E., Holzer T.E., Fl$\dot {{\rm a}}$ T., 1982, Space Sci.
Rev. 33, 161

\noindent   Lucy L.B., Solomon P.M., 1970, ApJ 159, 879

\noindent \hangindent=0.5 true cm MacGregor K.B., 1996, in {\it Solar and
Astrophysical MHD Flows}, ed. Tsinganos (Dordrecht: Kluwer), p. 333

\noindent   Mahesvaran M., Cassinelli J.P., 1988, ApJ 335, 931

\noindent   Massa D. et al., 1995, ApJ 452, L53

\noindent   Mathews W.G., Doane J.S., 1990, ApJ 352, 423

\noindent   McKee C.F., Begelman M.C., 1990, ApJ 358, 392

\noindent   Moffat A.F.J., 1994, Ap\&SS 221, 467

\noindent   
\hangindent=0.5 true cm Moffat A.F.J., 1996, in: Wolf-Rayet Stars in the
Framework of Stellar Evolution, $33^{rd}$ Li\`ege Int. Astroph. Coll.

\noindent   
\hangindent=0.5 true cm Moffat A.F.J., L\'epine S., Henriksen R.N., Robert
C., 1994, Ap\&SS 216, 55

\noindent   Moffat A.F.J., Robert C., 1994, ApJ 421,310

\noindent   
\hangindent=0.5 true cm Moss D., 1994, in Pulsation, Rotation and Mass Loss
in Early-Type Stars, Symp. IAU 126, p. 173

\noindent   Opher R., Pereira V.J.S., 1986, Astrophys. Lett. 25, 107

\noindent   Owocki S.P., Rybicki G.B., 1984, ApJ 284, 337

\noindent   Owocki S.P., Rybicki G.B., 1985, ApJ 299, 265

\noindent   Owocki S.P., Rybicki G.B., 1986, ApJ 309, 127

\noindent   Owocki S.P., 1994, Ap\&SS 221, 3

\noindent   Owocki S.P., Castor J.I., Rybicki G.B., 1988, ApJ 335, 914

\noindent   Owocki S.P., Rybicki G.B., 1991, ApJ 368, 261

\noindent   Parker E.N., 1953, ApJ 117, 431

\noindent   Poe C.H., Friend D.B., Cassinelli J.P., 1989, ApJ 337, 888

\noindent   Raymond J.C., Cox D.P., Smith B.W., 1976, ApJ 204, 290

\noindent   Rybicki G.B., Owocki S.P., 1990, ApJ 349, 274

\noindent   Robert C., 1994, Ap\&SS 221, 137

\noindent   Spitzer Jr L., 1978, Physics of Fully Ionized Gases, p. 148

\noindent   van der Hucht, 1992, A\&ARv 4, 123

\noindent   Velli M., 1993, A\&A 270, 304

\noindent   V$\ddot {{\rm o}}$lk H.J., Cesarsky C.J., 1982, Z. Naturforsch
37a, 809

\noindent 
\hangindent=0.5 true cm Willis A.J., 1991, in: WR Stars and Interrelations
with Other Massive Stars in Galaxies, K.A. van der Hucht and B. Hidayat
(eds.). Dordrecht: Kluwer, p. 265

\noindent   
\hangindent=0.5 true cm Wilson A.S., 1971, The Crab Nebula, IAU Symp. No.
46, eds. Davies R.D. \& Smith F.G., Reidel, Dordrecht

\newpage

\begin{figure}
\caption{
The module of the
heating/cooling processes, for $\log P_H=4.74$ ($P_H=5.5\times 10^4{\rm %
dyn/cm}^2$), $F_{nl}=F_t=F_A=10^4$ and $t=0.2$ (as in Fig. 2 - 5, where the
thermal instability process occurs for $T_L<T<T_H$ with $T_L=3\times
10^4\;K $ and $T_H=10^7\;K$).}
\psfig{file=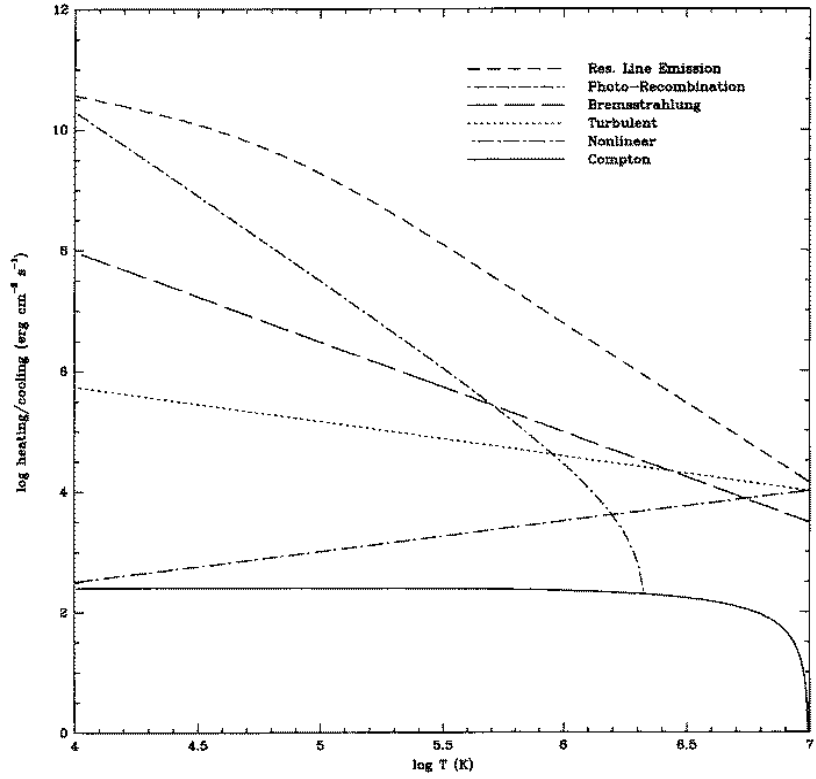,width=12.1truecm}
\end{figure}

\newpage

\begin{figure}
\caption{
The $\log T\times \log P$
diagram ($H(T,n)\approx 0$), with $F_A=10^2$: nonlinear and 
turbulent Alfvenic heatings.}
\psfig{file=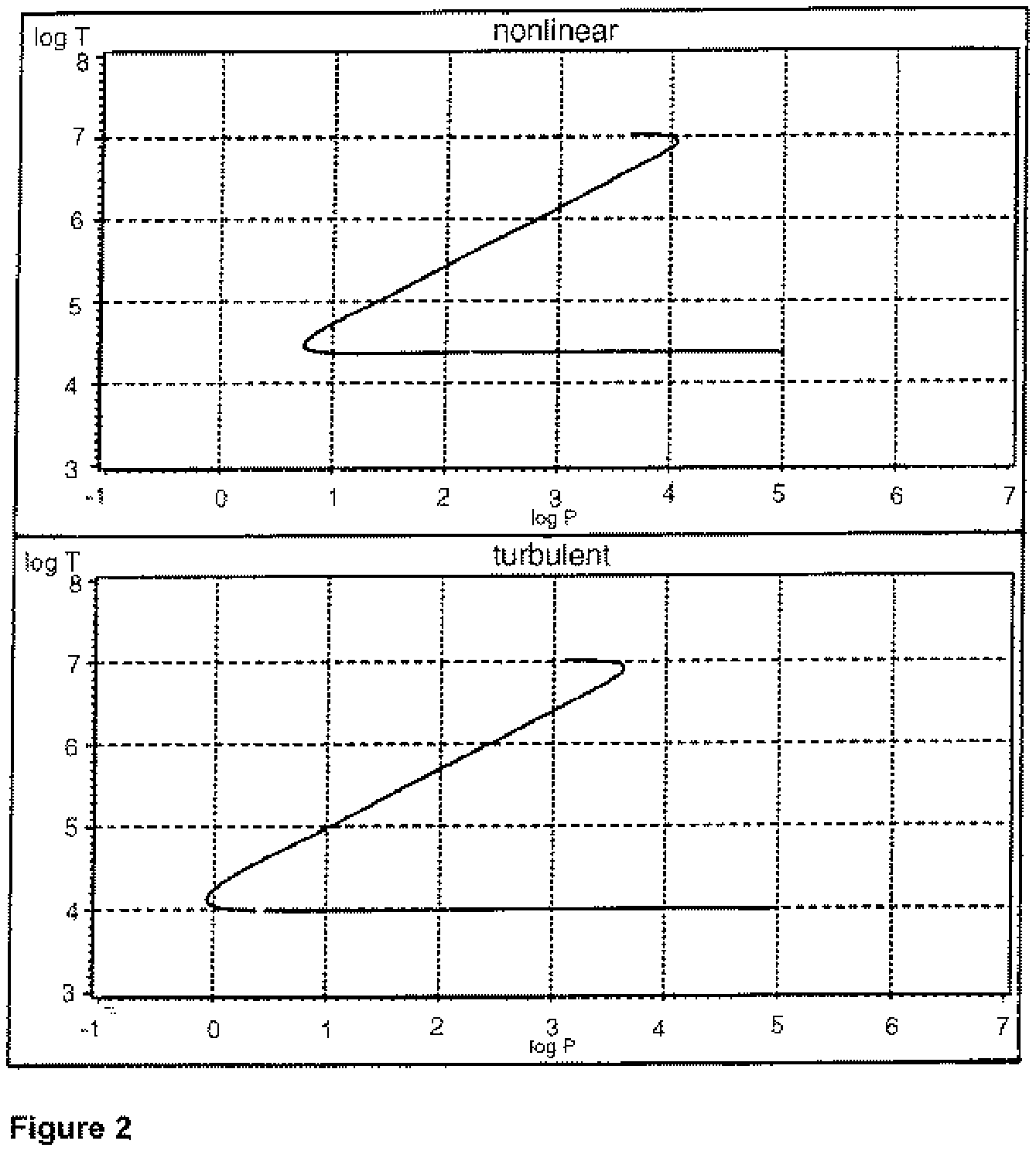,width=12.1truecm}
\end{figure}

\newpage

\begin{figure}
\caption{
The $\log T\times \log P$
diagram ($H(T,n)\approx 0$), with $F_A=5\times 10^3$: nonlinear and 
turbulent Alfvenic heatings.}
\psfig{file=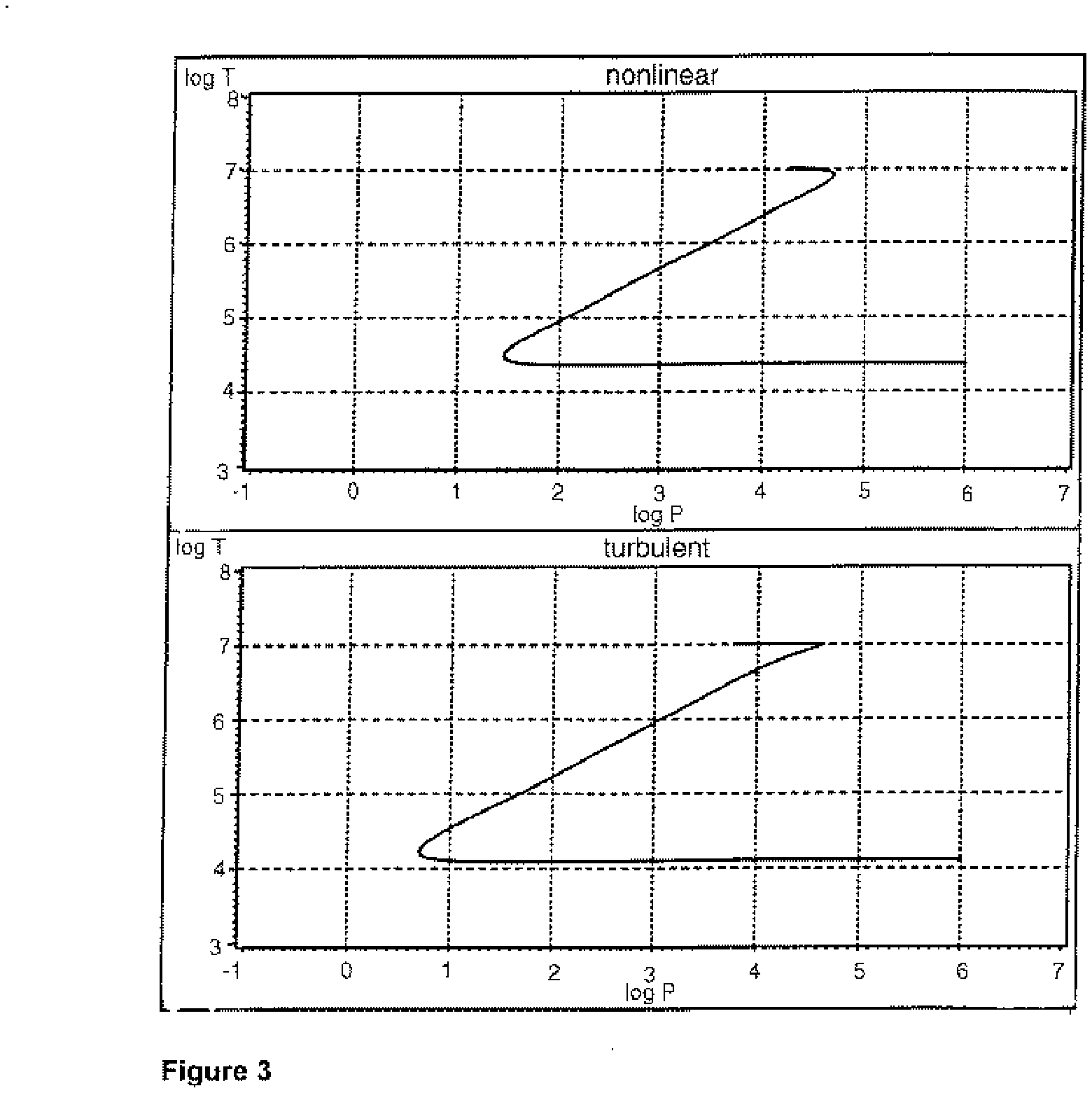,width=12.1truecm}
\end{figure}

\newpage

\begin{figure}
\caption{
The $\log T\times \log P$
diagram ($H(T,n)\approx 0$), with $F_A=10^4$: nonlinear and 
turbulent Alfvenic heatings.}
\psfig{file=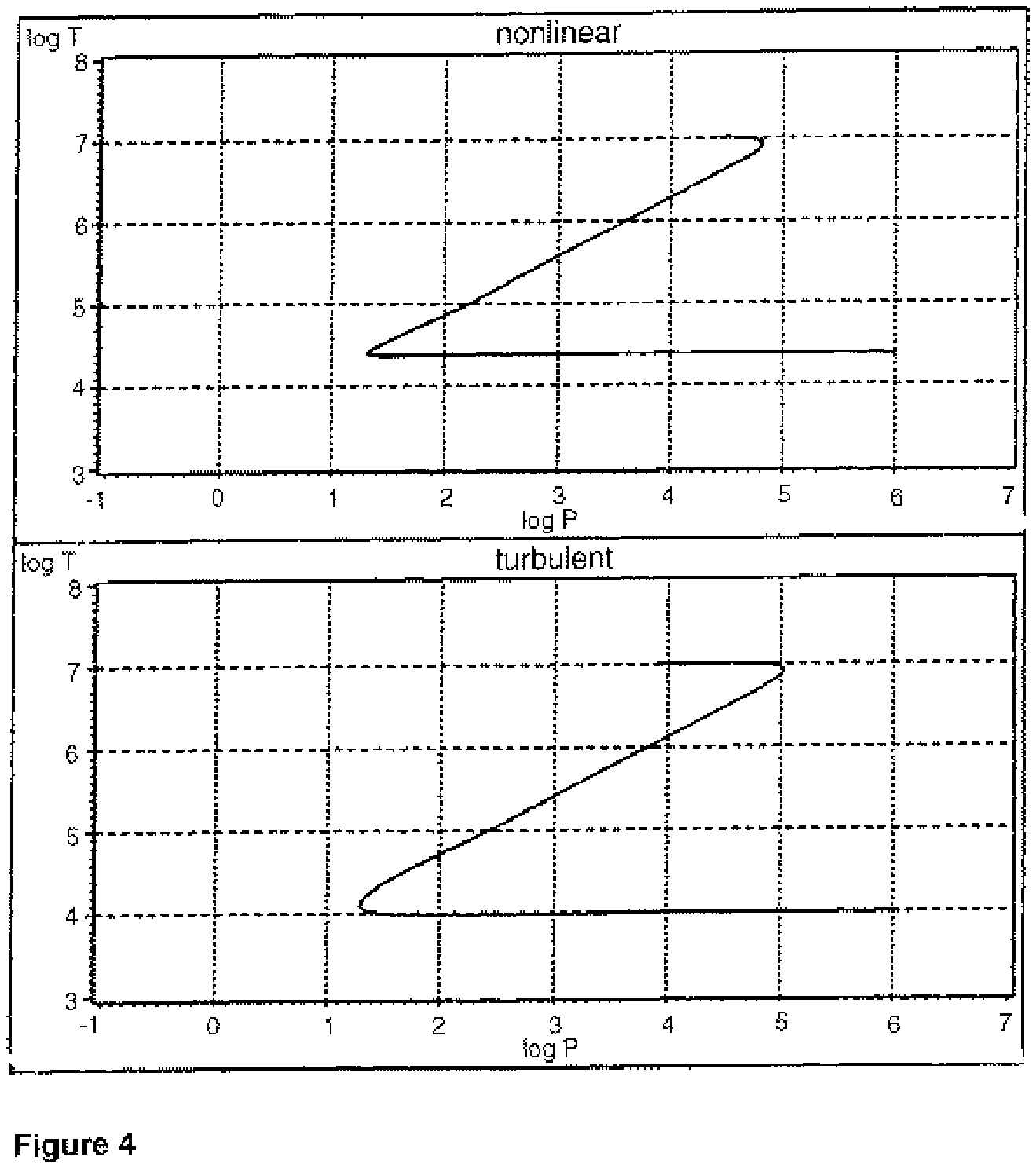,width=12.1truecm}
\end{figure}

\newpage

\begin{figure}
\caption{
The $\log T\times \log P$
diagram ($H(T,n)\approx 0$), with $F_A=10^5$: nonlinear and 
turbulent Alfvenic heatings.}
\psfig{file=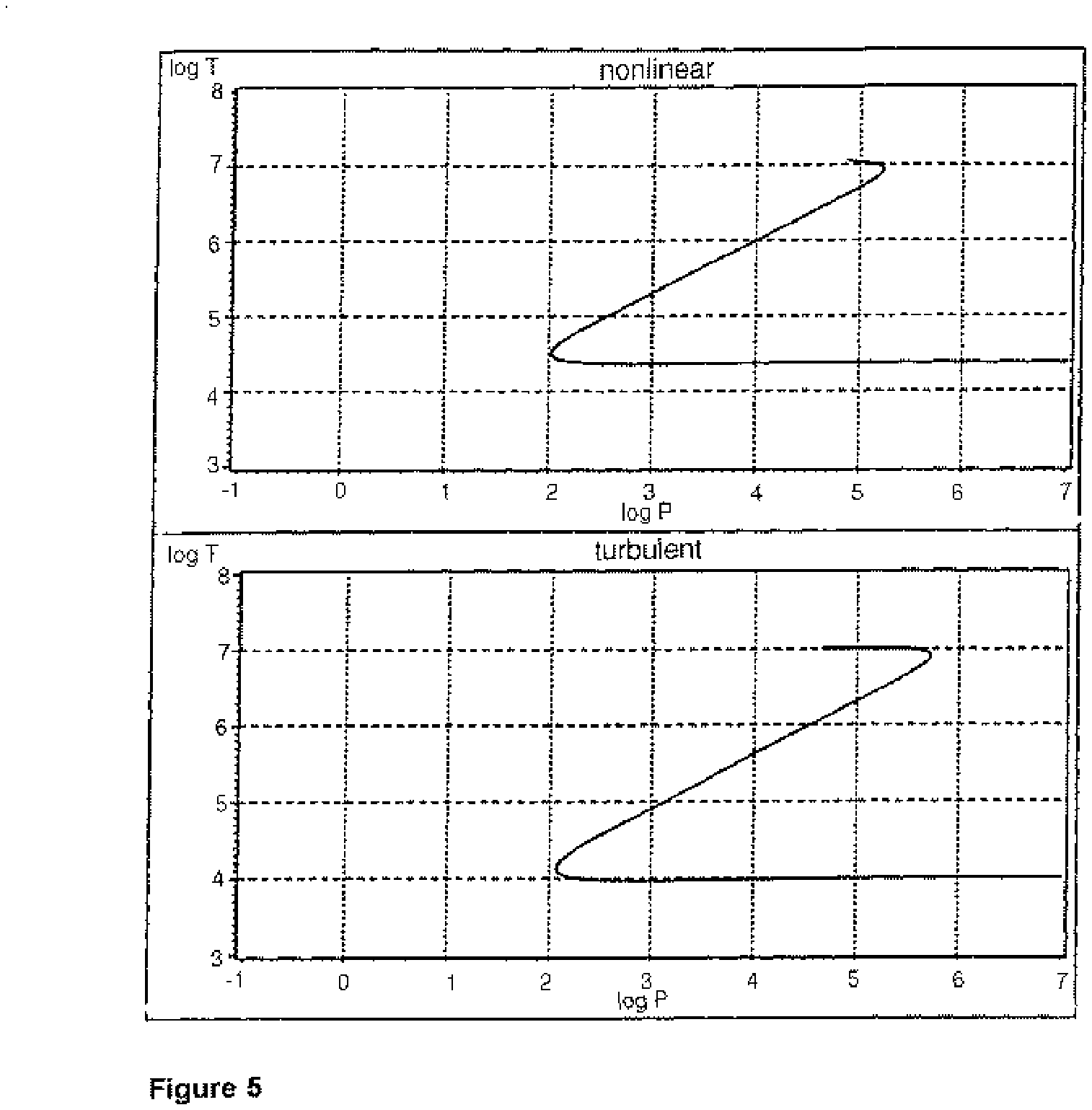,width=12.1truecm}
\end{figure}

\end{document}